\documentstyle[12pt,psfig]{article}
\begin{document}
\baselineskip=14pt
\parindent=0.25in
\abovedisplayskip=14pt
\belowdisplayskip=14pt
\parindent=0.5in
%
\renewcommand{\thefootnote}{\fnsymbol{footnote}}
\setcounter{footnote}{0}
\begin{center}
{\large\bf Hard Probes in High-energy Heavy-ion 
Collisions}\\
\vspace{0.5cm}
{\bf Xin-Nian Wang}\\
{\it Nuclear Science Division, Mailstop 70A-3307}\\
{\it Lawrence Berkeley National Laboratory}\\
{\it University of California, Berkeley, CA 94720 USA}\\
\end{center}
\vskip \baselineskip

\centerline{\bf ABSTRACT}
\centerline{\parbox[t]{5in}{
\baselineskip=12pt\small
Hard QCD processes in ultrarelativistic heavy-ion collisions
become increasingly relevant and they can be used as probes of
the dense matter formed during the violent scatterings. We will 
discuss how one can use these hard probes to study the properties
of the dense matter and the associated phenomenologies. In particular,
we study the effect of jet quenching due to medium-induced
energy loss on inclusive particle $p_T$ distributions and investigate 
how one can improve the measurement of parton energy loss in direct 
photon events.}}
 
\vskip 2\baselineskip

\baselineskip=14pt
\section{Introduction}

Quantum Chromodynamics (QCD) has been established as the underlying theory
of strong interactions. This theory is probably best demonstrated
in $e^+e^-$ annihilation processes in which a quark-antiquark
pair is produced and then fragmented into hadrons. In the earliest time
or at the shortest distance, one can use perturbative QCD(pQCD) to describe 
the interaction, i.e., parton radiation. Later on, the produced partons
will then combine with each via nonperturbative interaction and finally
hadronize into hadrons. Therefore, one can consider that there exists an 
interacting parton system during the prehadronization stage in $e^+e^-$
annihilation processes, which is, however, limited to a space-time
region characterized by the confinement scale $\Lambda_{\rm QCD}$.
The characteristic particle spectrum (in $p_T$ and rapidity) and the
ratios of produced particles should then be determined by the physics
of pQCD and nonperturbative hadronization. One seeks to produce a similar 
interacting parton system or a quark-gluon plasma but at a much larger 
scale of the order of a nucleus size in ultrarelativistic heavy-ion 
collisions. Therefore, one should study those experimental observables 
which are unique to the large size and long life-time of an interacting
partonic system as signals of a quark-gluon plasma.

        There are many proposed signals of a quark-gluon plasma \cite{BM95}.
Among them, hard probes associated with hard processes are especially
useful because they are produced in the earliest stage of the interaction
and their abilities to probe the dense matter are less complicated by
the hadronization physics. The merits of hard probes are even more
apparent at high energies because those processes are also dominating
the underlying collision dynamics which will determine the initial
conditions of the produced partonic system \cite{KE97,XNW97}. Study
of them will then enable us to probe the early parton dynamics and the
evolution of the quark-gluon plasma. In general, one can divide the
hard probes into two categories: thermal emission and particle suppression
by the medium. Particle production, like photon/dilepton and charm particles,
from thermal emission can be considered as the thermometers of the dense 
medium. Their background comes from the direct production in the initial 
collision processes. On the other hand, suppression of particles produced 
in the initial hard processes, like high-$p_T$ particles from jets 
and $J/\Psi$, can reveal evidences of the parton energy loss in
dense matter and the deconfinement of the partonic system. Thermal
production of these particles is expected to be negligible. Therefore,
in both cases, one needs to know the initial production rate accurately
enough. One advantage of these hard probes is that the initial production
rate can be calculated via pQCD, especially if we understand the slightly
nuclear modification one would expect to happen. In this talk, I will
only concentrate on high-$p_T$ particle suppression due to jet quenching.

Medium-induced radiative energy loss of a high-energy parton
traversing a dense QCD medium is interesting because
it depends sensitively on the density of the medium and thus can be 
used as a probe of the dense matter formed in ultrarelativistic heavy-ion
collisions. As recent studies demonstrated \cite{GW1,BDPS,BDMPS}, 
it is very important to take into account the destructive
interference effect in the calculation of radiation
spectrum from a fast parton induced by multiple scatterings.
The so-called Landau-Pomeranchuk-Migdal effect can lead
to very interesting, and sometimes nonintuitive results
for the energy loss of a fast parton in a QCD medium.
For example, Baier {\it et al}. recent showed \cite{BDMPS} that the 
energy loss per distance, $dE/dx$, is proportional to the total 
length that the parton has traveled. Because of the unique interference
effect, the parton somehow knows its history of propagation.
Another feature of the induced energy loss is that it depends
on the parton density of the medium it is traversing via
the final transverse momentum broadening that the parton receives
during its propagation. One can therefore determine the
parton density of the produced dense matter by measuring
the energy loss of a fast parton when it propagates
through the medium.

Unlike in the QED case, where one can measure directly the 
radiative energy loss of a fast electron, one cannot measure 
directly the energy loss of a fast leading parton in QCD.
Since a parton is normally studied via a jet, a cluster
of hadrons in the phase space, an identified jet can
contain particles both from the fragmentation of the
leading parton and from the radiated partons. If we neglect
the $p_T$ broadening effect, the total energy of the
jet should not change even if the leading parton
suffers radiative energy loss. What should be changed
by the energy loss are the particle distributions inside 
the jet or the fragmentation function and the jet profile.
Therefore, one can only measure parton energy loss indirectly
via the modification of the jet fragmentation function and
jet profile.

In principle, one can measure the parton energy loss by
directly measuring the fragmentation function and profile of
a jet with a determined transverse energy. However, because 
of the huge background and its fluctuation in high-energy
heavy-ion collisions \cite{WG90}, the conventional calorimetric study of 
a jet cannot determine the jet energy thus the energy 
loss very well.  In Ref.~\cite{WG92}, Gyulassy and I
proposed that single-particle
spectrum can be used to study the effect of jet energy loss,
since the suppression of large $E_T$ jets naturally leads to
the suppression of large $p_T$ particles. However, since the 
single-particle spectrum is a convolution of the jet cross
section and jet fragmentation function, the suppression of
produced particles with a given $p_T$ results from jet
quenching with a wide range of initial transverse energies.
One, therefore, cannot measure directly, from the single-particle
$p_T$ spectrum, the energy loss of
a jet with known initial transverse energy. Recently, Huang,
Sarcevic and I proposed to study the jet quenching by measuring
the $p_T$ distribution of charged hadrons in the opposite
direction of a tagged direct photon \cite{WHS}. Since a
direct photon in the central rapidity region ($y=0$) is
always accompanied by a jet in the opposite transverse
direction with roughly equal transverse energy, the $p_T$
distribution of charged hadrons in the opposite direction
of the tagged direct photon is directly related to the
jet fragmentation function with known initial energy.
One can thus directly measure the modification of the
jet fragmentation and then determine the energy loss
suffered by the leading parton.

In this talk, I will review the effects of energy loss
on single-particle distributions both in the normal central
$A+A$ collisions and in events with a tagged direct photon
with known transverse energy. I will discuss the energy
and $A$ dependences of the energy loss and jet quenching.
In the case of jet quenching in $\gamma +{\rm jet}$ events, 
the $E_T$ smearing of jet due to initial state radiation
will be included. The change of jet profile function in the
azimuthal angle due to $p_T$ broadening of the jet will
also be discussed. Finally, discussions will be given on
the feasibility of measuring the energy loss in $\gamma+{\rm jet}$
events at RHIC.

\section{Modified jet fragmentation functions}

Jet fragmentation functions have been studied extensively in $e^+e^-$,
$ep$ and $p\bar{p}$ collisions \cite{mattig}. These functions describe
the particle distributions in the fractional energy, $z=E_h/E_{jet}$, 
in the direction of a jet. The measured dependence of the fragmentation
functions on the momentum scale is shown to satisfy the QCD evolution
equations very well. We will use the parametrizations of the most
recent analysis \cite{bkk} in both $z$ and $Q^2$ for jet fragmentation 
functions $D^0_{h/a}(z,Q^2)$ to describe jet ($a$) fragmentation
into hadrons ($h$) in the  vacuum. 

In principle, one should study the modification of jet fragmentation
functions in a perturbative QCD calculation in which induced 
radiation of a propagating parton in a medium and 
Landau-Pomeranchuk-Migdal interference effect can be dynamically
taken into account. However, for the purpose of our current
study, we can use a phenomenological model to describe the
modification of the jet fragmentation function due to an
effective energy loss $dE/dx$ of the parton. In this model
we assume: (1) A quark-gluon plasma (QGP) is formed with 
a transverse size of the colliding nuclei, $R_A$. A parton 
with a reduced energy will only hadronize outside the deconfined 
phase and the fragmentation can be described as in $e^+e^-$ 
collisions. (2) The inelastic scattering mean-free-path
for the parton $a$ inside the QGP is $\lambda_a$. The radiative
energy loss per scattering is $\epsilon_a$. The energy
loss per distance is thus $dE_a/dx=\epsilon_a/\lambda_a$.
The probability for a parton to scatter $n$ times within
a distance $\Delta L$ is given by a Poisson distribution,
\begin{equation}
  P_a(n,\Delta L)=\frac{(\Delta L/\lambda_a)^n}{n!} 
  e ^{-\Delta L/\lambda_a} \; .
\end{equation}
We also assume that the mean-free-path of a gluon is half
that of a quark, and the energy loss $dE/dx$ is twice that
of a quark. (3) The emitted gluons, 
each carring energy $\epsilon_a$ on the average,
will also hadronize according to the fragmentation function 
with the minimum scale $Q_0^2= 2.0 $ GeV$^2$. We will also 
neglect the energy fluctuation given by the radiation spectrum
for the emitted gluons. Since the emitted gluons only produce
hadrons with very small fractional energy, the final modified
fragmentation functions in the moderately large $z$ region
are not very sensitive to the actual radiation spectrum and 
the scale dependence of the fragmentation functions for the 
emitted gluons.

We will consider the central rapidity region of high-energy
heavy-ion collisions. We assume that a parton with initial 
transverse energy $E_T$ will travel in the transverse direction 
in a cylindrical system. With the above assumptions, the modified 
fragmentation functions for a parton traveling a distance $\Delta L$
can be approximated as,
\begin{eqnarray}
  D_{h/a}(z,Q^2,\Delta L)& =&
  \frac{1}{C^a_N}\sum_{n=0}^NP_a(n,\Delta L)\frac{z^a_n}{z}D^0_{h/a}(z^a_n,Q^2)
  \nonumber \\
  &+&\langle n_a\rangle\frac{z'_a}{z}D^0_{h/g}(z'_a,Q_0^2), 
  \label{eq:frg1}
\end{eqnarray}
where $z^a_n=z/(1-n\epsilon_a/E_T)$, $z'_a=zE_T/\epsilon_a$ and
$C^a_N=\sum_{n=0}^N P_a(n)$. We limit the number of inelastic
scatterings to $N=E_T/\epsilon_a$ by energy conservation.
{}For large values of $N$, the average number of scatterings
within a distance $\Delta L$ is approximately 
$\langle n_a\rangle \approx \Delta L/\lambda_a$.
The first term corresponds to the fragmentation of the 
leading partons with reduced energy $E_T-n\epsilon_a$
and the second term comes from the emitted gluons each
having energy $\epsilon_a$ on the average.

\section{Energy loss and single-particle $p_T$ spectrum}

To calculate the $p_T$ distribution of particles from jet
fragmentation in a normal central heavy-ion collision, one 
simply convolutes the fragmentation functions with the jet 
cross section \cite{owens},
\begin{eqnarray}
  \frac{dN_{hard}^{AA}}{dyd^2p_T}&=&K\int d^2r\sum_{abcdh}
  \int_{x_{amin}}^1 dx_a \int_{x_{bmin}}^1 dx_b 
  f_{a/A}(x_a,Q^2,r)f_{b/A}(x_b,Q^2,r) \nonumber \\
  & & \frac{D_{h/c}(z_c,Q^2,\Delta L)}{\pi z_c}
  \frac{d\sigma}{d\hat{t}}(ab\rightarrow cd), \label{eq:nch}
\end{eqnarray}
where $z_c=x_T(e^y/x_a +e^{-y}/x_b)/2$, 
$x_{bmin}=x_ax_Te^{-y}/(2x_a-x_Te^y)$,
$x_{amin}=x_Te^y/(2-x_Te^{-y})$, and $x_T=2p_T/\sqrt{s}$.
The $K\approx 2$ factor accounts for higher order 
corrections \cite{xwke}. The parton distribution density in 
a nucleus, $f_{a/A}(x,Q^2,r)=t_A(r)S_{a/A}(x,r)f_{a/N}(x,Q^2)$, 
is assumed to be factorizable into the nuclear thickness 
function  $t_A(r)$ (with normalization $\int d^2r t_A(r)=A$),
parton distribution in a nucleon $f_{a/N}(x,Q^2)$ and the
parton shadowing factor $S_{a/A}(x,r)$ which we take the
parametrization used in HIJING model \cite{hijing}.
Neglecting the transverse expansion, the transverse
distance a parton produced at $(r,\phi)$ will travel is 
$\Delta L(r,\phi)=\sqrt{R_A^2-r^2(1-\cos^2\phi)}-r\cos\phi$.

\begin{figure}
\parbox[t]{2.8in}{\psfig{figure=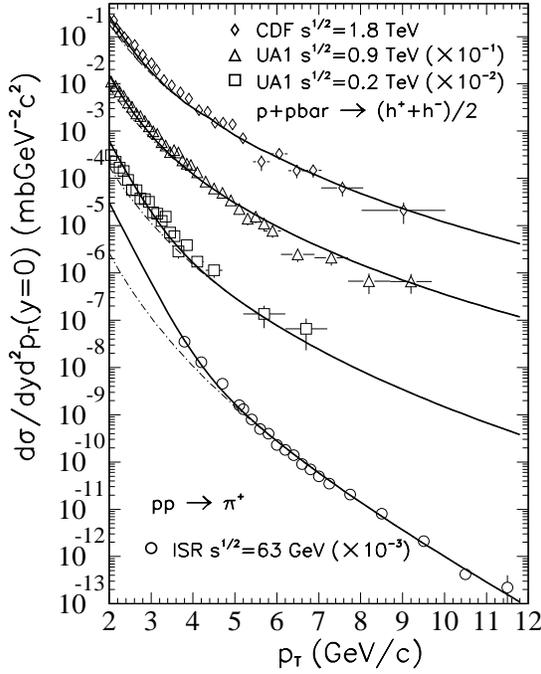,width=2.8in,height=3.5in}}
\parbox[b]{2in}{
\caption{ The charged particle $p_T$ spectra in $pp$ and $p\bar{p}$
collisions. The dot-dashed lines are from jet fragmentation only
and solid lines include also soft production parametrized in an
exponential form. The experimental data are from 
Ref.~\protect\cite{alper,ua1pt,cdfpt}.} \vspace{0.5in}
\label{figjet1}}
\end{figure}

We will use the MRS D$-\prime$ parametrization of the parton
distributions \cite{mrs} in a nucleon. The resultant $p_T$-spectra
of charged hadrons ($\pi^{\pm}, K^{\pm}$) for $pp$ and $p\bar{p}$
collisions are shown in Fig.~\ref{figjet1} together with the
experimental data \cite{alper,ua1pt,cdfpt} 
for $\sqrt{s}=63$, 200, 900 and 1800 GeV.
The calculations (dot-dashed line) from Eq.~(\ref{eq:nch}) 
with the jet fragmentation functions given by Ref.~\cite{bkk} 
agree with the  experimental data remarkably well, especially
at large $p_T$. However, the calculations are consistently
below the experimental data at low $p_T$, where we believe
particle production from soft processes is very important.
To account for particle production at small $p_T$, we introduce
a soft component to the particle spectra in an exponential form
whose parameters are fixed by total $dN/dy$ \cite{wang2}.
The total $p_T$ spectra including both soft and hard component
are shown in Fig.~\ref{figjet1} as solid lines.

\begin{figure}
\parbox[b]{2in}{
\caption{The ratio of charged particle $p_T$ spectrum in
central $Au+Au$ collisions at $\protect\sqrt{s}=200$ GeV over that of
$pp$ collisions, normalized by the total binary nucleon-nucleon
collisions in central $Au+Au$ collisions. The mean-free-path
of a quark inside the medium is assumed to be 1 fm.}\vspace{0.5in}
\label{figjet2}}
\parbox[t]{2.8in}{\psfig{figure=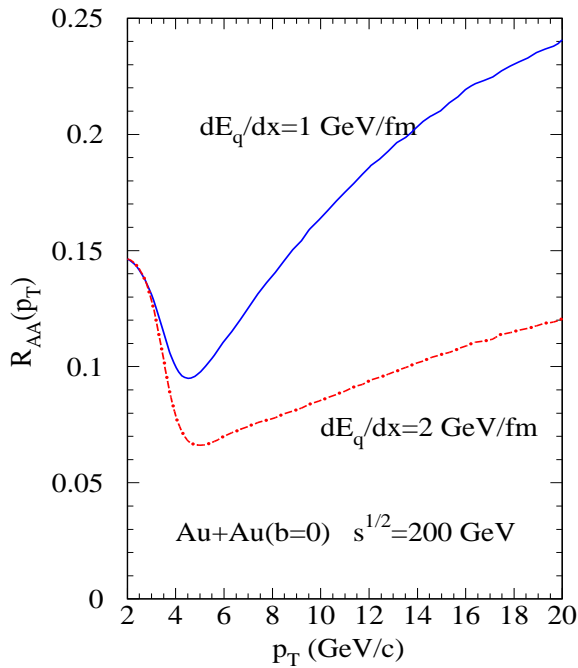,width=3in,height=3.5in}}
\end{figure}

To calculate the $p_T$ spectrum in $AA$ collisions, one has
to take into account both the parton shadowing effect and the
modification of the jet fragmentation functions due to parton
energy loss inside a medium. In addition, one has to know
the $A$ scaling of the soft particle spectrum in central
$AA$ collisions with respect to $pp$ collisions. Here we
simply assume a linear scaling as in a wounded nucleon
model. One can then calculate the ratio,
\begin{equation}
  R_{AA}(p_T)=\frac{dN_{AA}/dy/d^2p_T}{\sigma_{pp}T_{AA}(0)
    dN_{pp}/dy/d^2p_T} \; , \label{eq:ratio}
\end{equation}
between the spectrum in central $AA$ and $pp$ collisions. 
The ratio is normalized to the effective total number
of binary $pp$ collisions in a central $AA$ collision.
If none of the nuclear effects (shadowing and jet quenching) 
are taken into account, this ratio should  be unity at
large transverse momentum. Shown in Fig.~\ref{figjet2} are the
results for central $Au+Au$ collisions at the RHIC energy
with $dE/dx=1$, 2 GeV/fm, respectively. As we have
argued before, jet energy loss will result in the suppression
of high $p_T$ particles as compared to $pp$ collisions. Therefore,
the ratio at large $p_T$ in Fig.~\ref{figjet2} is smaller than one 
due to the energy loss suffered by the jet partons. It, however, 
increases with $p_T$ because of the constant energy loss ( or
even some weak energy dependent energy loss). At hypothetically
large $p_T$ when the total energy loss is negligible compared 
to the initial jet energy, the ratio should approach to one.

At small $p_T$, particles from soft interaction (or from 
hadronization of QGP) dominate. The ratio $R_{AA}(p_T)$
is very sensitive to the $A$-scaling behavior of the soft
particle production. Since we assumed a linear scaling
for the soft particle production, the ratio should approach
to $A/\sigma_{pp}T_{AA}(0)=0.149$ at small $p_T$ for
central $Au+Au$ at the RHIC energy, as shown in Fig.~\ref{figjet2}.

In this framework, one can also study the effect of energy
and $A$ dependence of the energy loss and the effect of
energy loss on particle production of different flavors \cite{wang2}.

\section{Jet quenching in $\gamma$+jet events}

Since hadron production at a fixed large $p_T$ comes from
fragmentation of jets with different transverse energies,
the suppression factor in Eq.~(\ref{eq:ratio}) only provides
the information about the effect of energy loss on jet fragmentation
at an averaged value of $z=p_T/E^{\rm jet}_T$.  In order to study
the modification of the fragmentation function due to energy loss,
one might in principle measure the inclusive $p_T$ spectrum in the
direction of a triggered jet. However, with the large background
and its fluctuation due to hadrons from many other minijets and
soft processes, the determination of the jet energy is almost
impossible. To overcome this difficulty, we proposed the study
of high $p_T$ particle spectrum in the opposite direction of
a tagged direct photon \cite{WHS}. Direct photons are always
accompanied by a jet in the opposite transverse direction. 
Even though taking into account of the initial state radiation, 
the average energy of the jet is approximately that of the
tagged photon. One can therefore relate the $p_T$ distribution
of hadrons in the opposite direction of a tagged photon to
the fragmentation function of a jet with known initial
energy and study the modification of the fragmentation function
due to parton energy loss.

\begin{figure}
\centerline{\psfig{figure=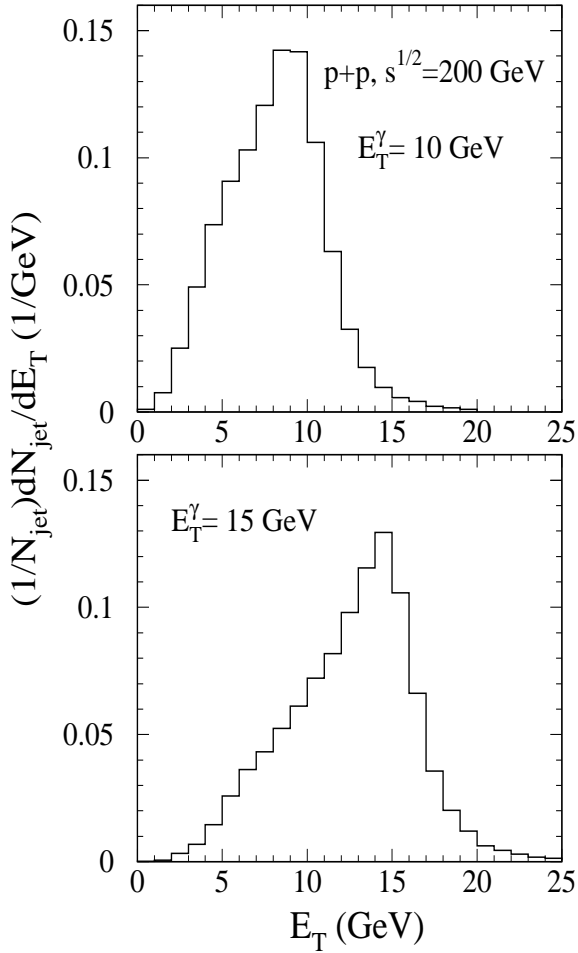,width=3in,height=5in}}
\caption{ The $E_T$ distributions of the jet, caused by the
initial state radiations, accompanying a triggered direct photon
with $E_T^{\gamma}=10$, 15 GeV, respectively at $\protect\sqrt{s}=200$ GeV.}
\label{figjet3}
\end{figure}

Let us select events which has a direct photon with transverse 
energy $E^{\gamma}_T$ in the central rapidity 
region, $|y|\leq \Delta y/2$, $\Delta y=1$. 
{}For sufficiently large $E_T^{\gamma}$ of the photon, the rapidity 
distribution of the associated jet is also centered around 
zero rapidity with a comparable width. If the azimuthal 
angle of the photon is $\phi_{\gamma}$ and 
$\bar{\phi}_{\gamma}=\phi_{\gamma}+\pi$, most of the hadrons
from the jet fragmentation will fall into the kinematic region,
$(|y|\leq \Delta y/2, |\phi-\bar{\phi}_{\gamma}|\leq \Delta\phi/2)$,
where one can take $\Delta\phi=2$ according to the jet profile
as measured in high-energy $p\bar{p}$ collisions \cite{ua1}.
Given the jet fragmentation functions $D_{h/a}(z)$, with $z$ 
the fractions of momenta of the jet carried by hadrons, one can 
calculate the differential $p_T$ distribution of hadrons from jet 
fragmentation in the kinematical region $(\Delta y,\Delta \phi)$,
\begin{equation}
  \frac{dN_{pp}^{\gamma-jet}}{dyd^2p_T}=
  \sum_{a,h}r_a(E_T^{\gamma})\int dE_T^{jet} g(E_T^{jet},E_T^{\gamma})
  \frac{D_{h/a}(p_T/E_T^{jet})}{p_T E_T^{jet}} 
  \frac{C(\Delta y,\Delta\phi)}{\Delta y\Delta\phi}, \label{eq:frg2}
\end{equation}
where $C(\Delta y,\Delta\phi)=\int_{|y|\leq \Delta y/2}dy
\int_{|\phi-\bar{\phi}_{\gamma}|
\leq \Delta\phi/2}d\phi f(y,\phi-\bar{\phi}_{\gamma})$
is an overall acceptance factor and $f(y,\phi)$ is the normalized
hadron profile around the jet axis. 
The summation is over both jet ($a$) and hadron species ($h$),
and $r_a(E_T^{\gamma})$ is the fractional production cross 
section of $a$-type jet associated with the direct photon.
$C(\Delta y,\Delta\phi)$ is the acceptance factor for finding 
the jet fragments in the given kinematic range.
We find $C(\Delta y,\Delta\phi)\approx 0.5$
at $\sqrt{s}=200$ GeV, independent of the photon energy $E_T^{\gamma}$,
using HIJING \cite{hijing} Monte Carlo simulations for 
the given kinematic cuts.  The normalized function,
\begin{equation}
  g(E_T^{jet},E_T^{\gamma})=\frac{1}{N_{\gamma-jet}}
  \frac{dN_{\gamma-jet}}{dE_T^{jet}}\; ,
\end{equation}
as shown in Fig.~\ref{figjet3}, is the $E_T$ distribution of the jet
with a given $E_T^{\gamma}$ of the tagged direct photon.
As we can see that the transverse energy of the jet has
a wide smearing around $E_T^{\gamma}$ due to the initial
state radiation associated with the hard processes. Because
of the rapidly decrease in $E_T$ of the cross section of 
direct photon production, the distribution is biased toward 
smaller $E_T^{jet}$ than $E_T^{\gamma}$. The average $E_T^{jet}$
is thus smaller than $E_T^{\gamma}$. Since one only triggers
a direct photon with a given  $E_T^{\gamma}$, one should
average over the $E_T$ smearing of the jet. Such a smearing
is important especially for hadrons with $p_T$ comparable or
larger than $E_T^{\gamma}$.

\begin{figure}
\centerline{\psfig{figure=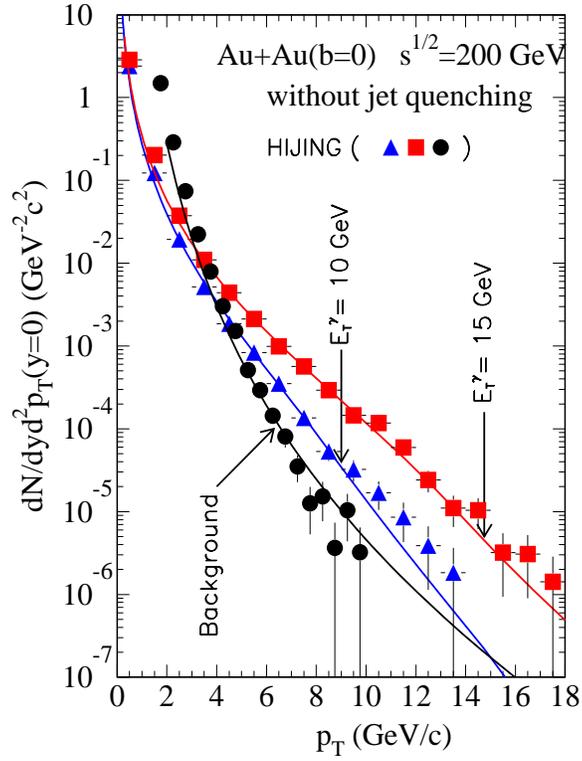,width=3in,height=4in}}
\caption{ The differential $p_T$ spectrum of charged particles
  from the fragmentation of a photon-tagged jet with 
  $E_T^{\gamma}=$10, 15 GeV
  and the underlying background in central $Au+Au$ collisions at
  $\protect\sqrt{s}=200$ GeV. The direct photon is restricted to 
  $|y|\leq \Delta y/2=0.5$. Charged particles are limited to the
  same rapidity range and in the opposite direction of the photon,
  $|\phi-\phi_{\gamma}-\pi|\leq \Delta\phi/2=1.0$. Solid lines
  are from the jet fragmentation function and points are HIJING
  simulations of 10K events. Parton energy loss is not included yet.}
\label{figjet4}
\end{figure}

If we define the inclusive fragmentation function associated
with a direct photon as,
\begin{equation}
  D^{\gamma}(z,E_T^{\gamma})=\sum_{ah}r_a(E_T^{\gamma})
  \int dE_T^{jet} g(E_T^{jet},E_T^{\gamma})\frac{E_T^{\gamma}}{E_T^{jet}}
  D_{h/a}(z\frac{E_T^{\gamma}}{E_T^{jet}})\; ,
\end{equation}
we can rewrite the $p_T$ spectrum [Eq.~(\ref{eq:frg2})] in the 
opposite direction of a tagged photon as
\begin{equation}
  \frac{dN_{pp}^{\gamma-jet}}{dyd^2p_T}=
  \frac{D^{\gamma}(p_T/E_T^{\gamma}, E_T^{\gamma})}{p_T E_T^{\gamma}} 
  \frac{C(\Delta y,\Delta\phi)}{\Delta y\Delta\phi}\; . \label{eq:frg3}
\end{equation}
Using this equation, one can extract the inclusive jet fragmentation
function, $D^{\gamma}(z,E_T^{\gamma})$, from the measured spectrum.
Shown in Fig.~\ref{figjet4} are the calculated $p_T$ distributions
from the fragmentation of photon-tagged jets with 
$E_T^{\gamma}=10$, 15 GeV and the underlying background from
the rest of a central $Au+Au$ collisions at the RHIC energy.
The points are HIJING simulations of 10K events and solid lines
are numerical results from Eqs.~(\ref{eq:nch}) and (\ref{eq:frg2})
with the fragmentation functions given by the parametrization
of $e^+e^-$ data \cite{bkk}. The effect of parton energy loss is
not included yet. As we can see, the spectra from jet 
fragmentation are significantly higher than the background
at large transverse momenta. One can therefore easily
extract the fragmentation function from the experimental data
without much statistical errors from the subtraction of
the background. One also notice that there are significant
number of particles with $p_T$ larger than the triggered
photon, $E_T^{\gamma}$, because of the $E_T$ smearing of
the jet caused by initial state radiations.

Consider parton energy loss in central $AA$ collisions,
we model the jet fragmentation functions as given by
Eq.~(\ref{eq:frg1}). Including the $E_T$ smearing
and averaging over the $\gamma$-jet production position
in the transverse direction, the inclusive fragmentation 
function of a photon-tagged jet is,
\begin{equation}
  D^{\gamma}_{AA}(z)=\int \frac{d^2r t^2_A(r)}{T_{AA}(0)}\sum_{ah} 
  r_a(E_T^{\gamma})
  \int dE_T^{jet} g(E_T^{jet},E_T^{\gamma})\frac{E_T^{\gamma}}{E_T^{jet}}
  D_{h/a}(z\frac{E_T^{\gamma}}{E_T^{jet}},\Delta L) \; ,
\end{equation}
where $T_{AA}(0)=\int d^2r t^2_A(r)$ is the overlap function of
$AA$ collisions at zero impact-parameter. We assume that jet 
production rate is proportional to the number of binary 
nucleon-nucleon collisions.

\begin{figure}
\centerline{\psfig{figure=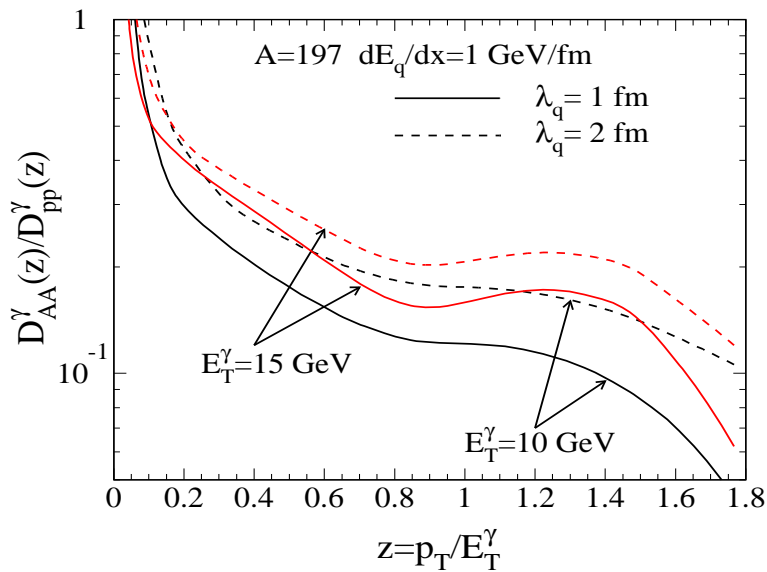,width=4in,height=3in}}
\caption{Ratio of the inclusive fragmentation function of
a photon-tagged jet with and without energy loss in central
$Au+Au$ collisions for a fixed $dE_q/dx=1$ GeV/fm.}
\label{figjet5}
\end{figure}

Shown in Fig.~\ref{figjet5} are the ratios of the inclusive
fragmentation function in a central $Au+Au$ collisions
with energy loss $dE_q/dx=1$ GeV/fm, over the ones in
$pp$ collisions without energy loss. The enhancement of soft
particle production due to induced emissions is important
only at small fractional energy $z$. The fragmentation
function is suppressed for large and intermediate $z$
due to parton energy loss. For a fixed $dE/dx$, the
suppression becomes less as $E_T^{\gamma}$ increases.
The optimal case is when the average total energy loss
is significant as compared to the initial jet energy,
and yet the $p_T$ spectrum from jet fragmentation is
still much larger than the underlying background.
Notice that we now define $z$ as the hadron's fractional 
energy of the triggered photon. Because of the $E_T$ smearing
of the jet caused by initial state radiations, hadrons can
have $p_T$ larger than $E_T^{\gamma}$. Therefore, the effective
inclusive jet fragmentation function does not vanish at
$z=p_T/E_T^{\gamma} > 1$.

\begin{figure}
\centerline{\psfig{figure=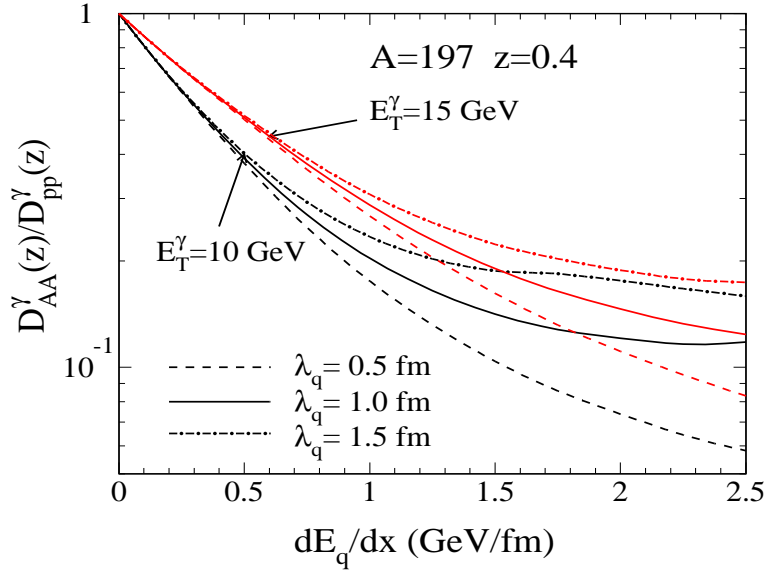,width=4in,height=3in}}
\caption{Ratio of the inclusive fragmentation function of
a photon-tagged jet with and without energy loss in central
$Au+Au$ collisions at $z=0.4$ as a function of $dE_q/dx$.}
\label{figjet6}
\end{figure}

As compared to our earlier results \cite{WHS} where
we did not take into account the $E_T$ smearing of the
jet, the modification of the averaged fragmentation
function due to energy loss is quite sensitive to
the value of the mean-free-path for $dE_q/dx=$ 1 GeV.
To study the sensitivity of the suppression to the
energy loss, we plot in Fig.~\ref{figjet6} the same
ratio at a fixed value of $z=0.4$ as functions of
$dE_q/dx$.  The ratio in general decreases with $dE_q/dx$
as more large $p_T$ particles are suppressed when
leading partons loss more energy. For small values
of $dE_q/dx$, the suppression factor is more or less
independent of the mean-free-path. However, for large values
of $dE_q/dx \geq 1$ GeV/fm, the ratio is sensitive to
the mean-free-path. One thus needs additional information
or a global fit to determine both the energy loss $dE/dx$ 
and the mean-free-path from the experimental data.

\begin{figure}
\centerline{\psfig{figure=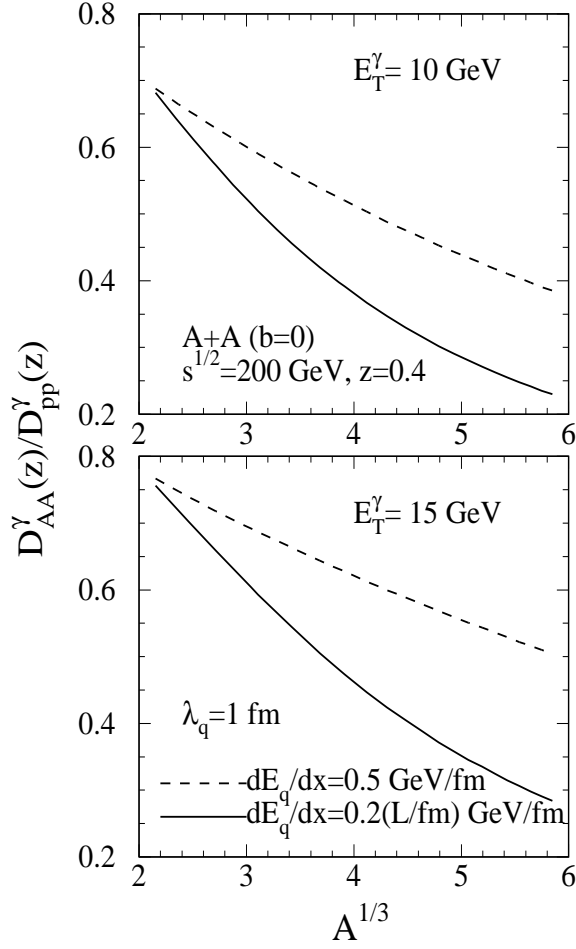,width=3in,height=5in}}
\caption{Ratio of the inclusive fragmentation function of
a photon-tagged jet with and without energy loss in central
$Au+Au$ collisions at $z=0.4$ as a function of $A^{1/3}$.
A constant energy loss, $dE_q/dx=0.5$ GeV/fm (dashed lines)
and $dE_q/dx=0.2 (L/{\rm fm})$ GeV/fm (solid lines), which is linear in
the total distance the jet has traveled, are assumed.}
\label{figjet7}
\end{figure}

As we discussed in the introduction, recent studies \cite{BDMPS}
of energy loss in a dense medium indicate that
the energy loss per distance $dE/dx$ might be proportional
to the total distance that the parton has traveled since
is was produced. One way to test this experimentally
is to study the suppression factor at any given $z$
value for different nucleus-nucleus collisions or for different
centrality (impact parameter). Shown in Fig.~\ref{figjet7},
are the suppression factor for the jet fragmentation function
at $z=0.4$ as functions of $A^{1/3}$. In one case (dashed lines),
we assume a constant energy loss $dE/dx$=0.5 GeV/fm. The 
suppression factor decreases almost linearly with $A^{1/3}$.
In another case (solid lines), we assume $dE_q/dx=0.2 (L/{\rm fm})$ GeV/fm.
The average distance a parton travels in a cylindrical system
with transverse size $R_A$ is $\langle L\rangle =0.905 R_A$.
We assume $R_A=1.2 A^{1/3}$ fm. We choose the coefficient 
in $dE_q/dx$ such that it roughly equals to 0.5 GeV/fm for
$A=20$. As we can see, the suppression factor for a 
distance-dependent $dE/dx$ decreases faster than
the one with constant $dE/dx$. Unfortunately, we have
not found a unique way to extract the average total 
energy loss so that we can show it is proportional
to $A^{2/3}$ for the distance-dependent $dE/dx$.

\section{$p_T$ broadening and jet profile}

In the above calculation, we have assumed that the jet profile
in the opposite direction of the tagged photon remains the
same in $AA$ collisions, since we used the same acceptance
factor $C(\Delta y, \Delta\phi)$. However, due to multiple
scatterings suffered by the leading parton, the final jet
must acquire additional acoplanarity with respect to its
original transverse direction. Such a change to the jet
profile could affect the acceptance factor, which will be
an overall factor to the measured jet fragmentation function
if we assume the jet profile to be the same for particles
with different fractional energies. 

\begin{figure}
\centerline{\psfig{figure=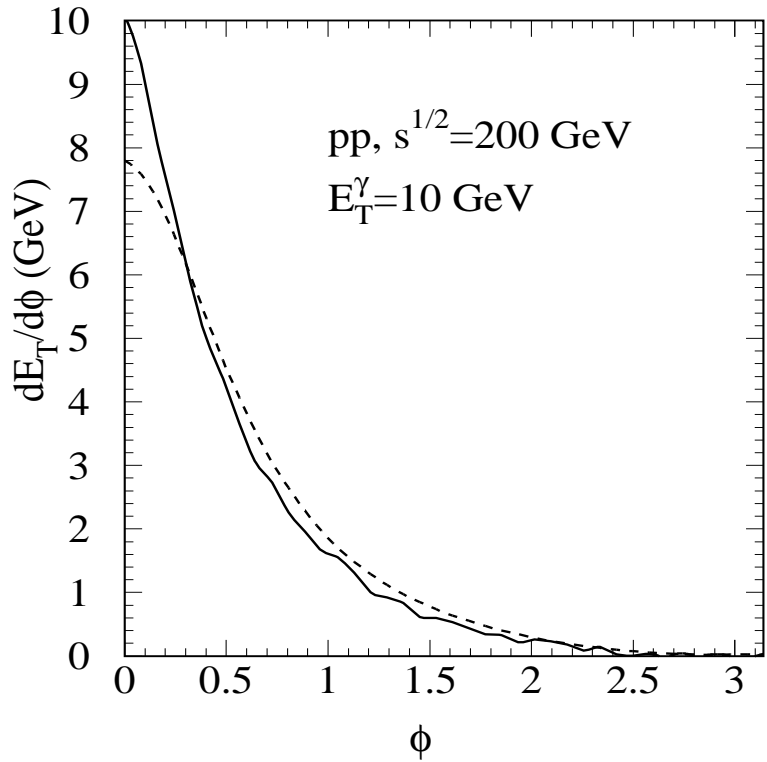,width=4in,height=4in}}
\caption{The effective profile function of the photon-tagged
jet, defined as the $E_T$ distribution in azimuthal angle
respect to the opposite direction of a photon with $E_T^{\gamma}=10$ GeV. 
The dashed line assumes a Gaussian form of $p_T$ broadening for the jet
with an average $\Delta p_T^2=4$ (GeV/$c$)$^2$.}
\label{figjet8}
\end{figure}

To demonstrate this, we plot in Fig.~\ref{figjet8} as the
solid line the azimuthal angle distribution of $E_T$ (within $|y|<0.5$) 
with respect to the opposite direction of the tagged photon
with $E_T^{\gamma}=10$ GeV. We have subtracted the background 
so that $dE_T/d\phi=0$ at $\phi=\pi$. The profile distribution 
includes both the intrinsic distribution from jet fragmentation 
and the effect of initial state radiation. The acceptance factor 
is simply the fractional area within $|\phi|<\Delta\phi/2$ region.
The $p_T$ broadening of jets due to multiple scatterings
will broaden the profile function. Shown as the dashed line
is the profile function for an average $\Delta p_T^2=4$ (GeV/$c$)$^2$
with a Gaussian distribution. It is clear that with a modest
value of the $p_T$ broadening, the acceptance factor only
changes by around 10\%\cite{WH2}.

In addition, since the change of the jet profile function is related
to the average $p_T$ broadening, one can combine the measurement
with the measured energy loss to verify the relationship between
$dE/dx$ and $\Delta p_T^2$ as suggested by recent theoretical
studies \cite{BDMPS}.

\section{Discussions}

To have a feeling of the experimental feasibility of the
proposed $\gamma-$jet measurement, we list in Table~1
the number of $\gamma-$jet events per year per unit rapidity and
unit (GeV) $E_T$. We assume a central $Au+Au$
cross section of 125 mb with impact-parameters $b<2$ fm.
We have taken the designed luminosity of ${\cal L}=2\times 10^{26}$
cm$^{-2}$s$^{-1}$ with 100 operation days per year.
As we can see, the rate for $E_T^{\gamma}=15$, 20 GeV is too
small to give any statistically significant measurement of
the fragmentation function and its modification in $AA$
collisions. If one can increase the luminosity by a factor
of 10, the numbers of events for both $E_T^{\gamma}=$10 and 15 GeV
are significant enough for a reasonable determination of
the fragmentation function of the photon-tagged jets.

\begin{table}
\begin{center}
\begin{tabular}{|l|llll|} \hline
$E_T^{\gamma}$ (GeV) & 7 & 10 & 15 & 20 \\ \hline
$dN^{\gamma-jet}/dydE_T/$year & 20500 & 3500 & 400 & 70 \\ \hline
\end{tabular}
\caption{Rate of direct photon production in central $Au+Au$
  collisions at $\protect\sqrt{s}=200$ GeV, with luminosity 
  ${\cal L}=2\times 10^{26}$ cm$^{-2}$s$^{-1}$ and 100 operation
  day per year.}
\end{center}
\end{table}

\begin{figure}
\centerline{\psfig{figure=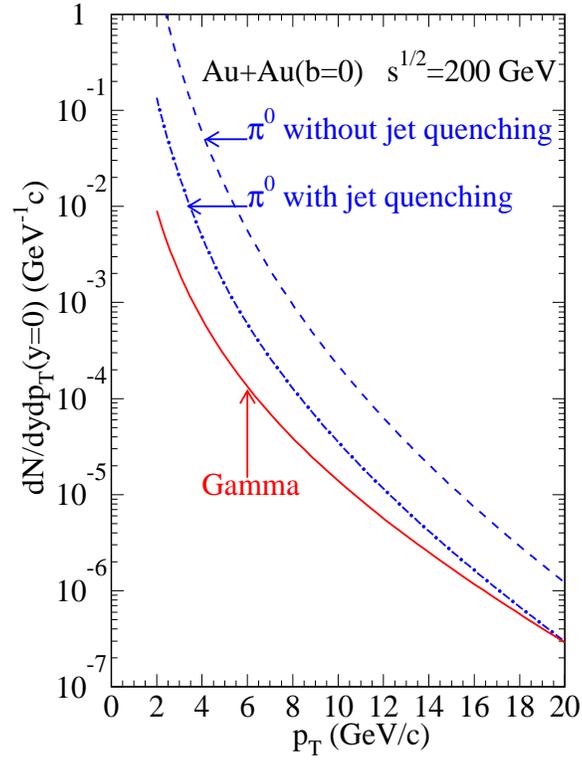,width=3in,height=4in}}
\caption{The inclusive $p_T$ distribution for direct photons as
compared to that of $\pi^0$'s with and without parton energy
loss in central $Au+Au$ collisions at $\protect\sqrt{s}=200$ GeV.
$dE_q/dx=1$ GeV/fm and mean-free-path $\lambda_q=1$ fm are assumed.}
\label{figjet9}
\end{figure}

Given increased luminosity and enough number of events, one still
has to overcome the large background of $\pi^0$'s to identify
the direct photons. Plotted in Fig.~\ref{figjet9}, are the production
rates of direct photons (solid line) and $\pi^0$'s (dashed and
dot-dashed lines). We can see that without parton energy loss,
$\pi^0$ production rate is about 20 times larger than the
direct photons at $p_T=10$ GeV/$c$. Fortunately, jet quenching 
due to parton energy loss can significantly reduce $\pi^0$
rate at large $p_T$ as shown by the dot-dashed line. However,
one still has to face $\pi^0$'s about 4 times higher than
the direct photons at $p_T=10$ GeV/$c$. At larger $p_T$,
the situation improves, but one loses the production rate.
Since the isolation cut method normally employed in $pp$ collisions
to reduce the background to direct photons does not work anymore, the
only way one can identify them has to be through improvement of
detector hardwares.

\section*{Acknowledgements}
Some of the work in this talk was done in collaboration with Z.~Huang
and I. Sarcevic. This work was supported by the Director, Office of Energy
Research, Office of High Energy and Nuclear Physics, Divisions of 
Nuclear Physics, of the U.S. Department of Energy under Contract No.\
DE-AC03-76SF00098 and DE-FG03-93ER40792.

\end{document}